\documentstyle[prb,aps,preprint,epsfig,amssymb,amsbsy,amsmath,multirow]{revtex}
\newcommand{\SMin}{S_{\mbox{\scriptsize min}}}
\newcommand{\op}[1]{%
    \fontdimen12\textfont3=2pt\fontdimen12\scriptfont3=1.4pt%
    \!\null\mathop{\vphantom{#1}\smash{#1}}\limits_{\sim}\null\!}

\newcommand{\vek}[1]{{\!\vec{\,#1}}}
\newcommand{\fmref}[1]{(\protect\ref{#1})}
\newcommand{\xref}[1]{\protect\ref{#1}}
\newcommand{\figref}[1]{Fig.~\protect\ref{#1}}
\def\bra#1{\langle \, {#1} \, | \,}
\def\ket#1{\, | \, {#1} \, \rangle}
\def\half{{\frac{1}{2}}}
\newcommand{\tr}{\mbox{tr}}
\newcommand{\pp}[2]{\frac{\partial \, {#1}}{\partial \, {#2}}\;}

\newcommand{\changed}[1]{#1}
\begin{document}
\tightenlines
\draft

\title{Ground state properties of
antiferromagnetic\\ Heisenberg spin rings}
\author{K. B\"arwinkel, H.-J. Schmidt, %
J. Schnack\cite{correspondence}} 
\address{Universit\"at Osnabr\"uck, Fachbereich Physik \\  
         Barbarastr. 7, 49069 Osnabr\"uck, Germany}

\date{\today}
\maketitle

\begin{abstract}
Exact ground state properties of antiferromagnetic Heisenberg
spin rings with isotropic next neighbour interaction are
presented for various numbers of spin sites and spin quantum
numbers. Earlier work by Peierls, Marshall, Lieb, Schultz and
Mattis focused on bipartite lattices and is not applicable to
rings with an odd number of spins. With the help of exact
diagonalization methods we find a more
general systematic behaviour which for instance relates the
number of spin sites and the individual spin quantum numbers to
the degeneracy of the ground state.  These numerical findings
all comply with rigorous proofs in the cases where a general
analysis could be carried out.  Therefore it can be plausibly
conjectured that the ascertained properties hold for ground
states of arbitrary antiferromagnetic Heisenberg spin rings.
\changed{These general rules help to explain the low temperature
behaviour of recently synthesized spin rings.}

\end{abstract}

\widetext
%%%%%%%%%%%%%%%%%%%%%%%%%%%%%%%%%%%%%%%%%%%%%%%%%%%%%%%%%%%%%%%%%%%%%%%%
\section{Introduction and summary}

Synthesized molecules containing relatively small numbers of
paramagnetic ions and their magnetic properties are of great
current interest
\cite{TDP94,GCR94,FST96,LGB97}. Some of them appear as rings of 
localized single-particle magnetic moments which are adequately
described by the Heisenberg model
\cite{BeG90,DGP93,CCF96,PDK97,WSK99} with \changed{-- in many
cases -- isotropic antiferromagnetic coupling.}

In this article properties of the ground state for
antiferromagnetically (AF) \changed{coupled spin rings} will be
presented. \changed{Exact diagonalization} methods
\cite{BeG90,Kou97,Kou98,BSS99,Wal99} make it possible to investigate
small spin rings for various numbers $N$ of spin sites and spin
quantum numbers $s$.  The obvious symmetries allow to decompose
the Hilbert space ${\mathcal H}$ into a set of mutually
orthogonal subspaces ${\mathcal H}(S,M,k)$ according to the
quantum numbers of the total spin $S$, the total magnetic
quantum number $M$ and the shift quantum number $k$ of the
cyclic shift operator.

In view of our exact numerical results (table \xref{T-2-1}),
several conjectures on general properties of the
antiferromagnetic ground state suggest themselves. The two most
important ones are:
\begin{itemize}
\item If $N\!\cdot\!s$ is integer, then the ground state is non-degenerate.
\item If $N\!\cdot\!s$ is half integer, then the ground state is fourfold degenerate.
\end{itemize}
These findings exceed those derived from the theorem of Lieb,
Schultz and Mattis \cite{LSM61,LiM62} and establish rules also
for spin rings of an odd number of spin sites.  Though, for the
time being, rigorous proofs or refutations of our statements in
the undecided cases remain a challenge, the numerical results
and partial proofs so far provide a strong evidence of general
validity. 

\changed{The presented general ground state properties are very
helpful in order to understand the low temperature behaviour of
small spin rings, since only a few states dominate the
properties at low temperature because of the finite system
size. In addition the found general rules allow statements about
spin rings of larger size whose Hamilton operator cannot be
diagonalized any longer. As an example the behaviour of the
zero-field susceptibility is discussed. Our results also open a
new view on frustration as they show that there are systems
which one would like to call frustrated, but which do not
possess the typical ground state degeneracy\cite{Cho86}.}

\section{Observations}

The Hamilton operator of the Heisenberg model with
antiferromagnetic, isotropic next-neighbor interaction between
spins of equal spin quantum number $s$ reads
%--------------------------------------------------------
\begin{eqnarray}
\label{E-2-1}
\op{H}
&=&
-
2\,J\,
\sum_{x=1}^N\;
\op{\vek{s}}(x) \cdot \op{\vek{s}}(x+1)
\ ,\quad \forall x: s(x)=s
\ ,\quad J < 0
\ ,
\end{eqnarray}
%--------------------------------------------------------
the spin sites being enumerated by $x$ modulo $N$.
This Hamilton operator commutes with the total spin
$\op{\vek{S}}$ and its three-component $\op{S}^3$. In addition
$\op{H}$ is invariant under cyclic shifts. These are represented
by a unitary cyclic shift operator $\op{T}$ defined by its
action on the product basis \fmref{E-2-2}
%--------------------------------------------------------
\begin{eqnarray}
\label{E-2-3}
\op{T}\,
\ket{m_1, \dots, m_{N-1}, m_N}
=
\ket{m_N, m_1, \dots, m_{N-1}}
\ ,
\end{eqnarray}
%--------------------------------------------------------
where the product basis of single-particle eigenstates of
all $\op{s}^3(x)$ obeys
%--------------------------------------------------------
\begin{eqnarray}
\label{E-2-2}
\op{s}^3(x)\,
\ket{m_1, \dots, m_x, \dots, m_N}
=
m_x\,
\ket{m_1, \dots, m_x, \dots, m_N}
\ .
\end{eqnarray}
%--------------------------------------------------------
For a shorter notation the basis elements are sometimes
abbreviated as
%--------------------------------------------------------
\begin{eqnarray}
\label{E-2-0}
\ket{\bold{m}}
:=
\ket{m_1, \dots, m_x, \dots, m_N}
\ .
\end{eqnarray}
%--------------------------------------------------------
The eigenvalues of $\op{T}$ are the $N$-th roots of unity,
%--------------------------------------------------------
\begin{eqnarray}
\label{E-2-4}
z
=
\exp\left\{
-i \frac{2\pi k}{N} 
\right\}
\ ,
\end{eqnarray}
%--------------------------------------------------------
where $k$ will be called shift quantum number, and $k$
can assume values $k=0,\dots, N-1$ modulo $N$.  
For every eigenstate of $\op{H}$ and $\op{T}$ with $k\ne 0$ and
$k\ne N/2$ 
there exists another eigenstate with translational quantum
number $N-k$ with the same energy eigenvalue, which is just the
result of invariance under complex conjugation.
For later use we define a ``cycle" as the linear span of all
product states which result from a given product state by
multiple repetition of the cyclic shift operator $\op{T}$
\cite{BSS99}. 

The Hamilton operator remains also invariant under spin flips
provided by the operator $\op{C}$, which turns all $m_x$ into
$-m_x$. In addition $\op{C}$ commutes with $\op{\vek{S}}^2$, but
anti-commutes with $\op{S}^3$. The eigenvalues of $\op{C}$ are
$\pi=\pm 1$.

\changed{Exact diagonalization} methods \cite{BSS99}
enable us to evaluate eigenvalues and eigenvectors of $\op{H}$
for small spin rings of various numbers $N$ of spin sites and
spin quantum numbers $s$. Our results for ground state
properties are summarized in table \xref{T-2-1}.  
\changed{For spin $s=1/2$ they are consistent with
Ref. \cite{FLM91} and for spin $s=1$ and even $N$ with
Ref. \cite{BoJ83}.}
Without exception we find:
\begin{enumerate}
\item The ground state belongs to the subspace ${\mathcal H}(S)$
with the smallest possible total spin quantum number $S$;
this is either $S=0$ for $N\!\cdot\!s$ integer, then the total magnetic quantum
number $M$ is also zero, or $S=1/2$ for $N\!\cdot\!s$ half integer, then
$M=\pm 1/2$.
\item The restricted ground state within a subspace 
of constant total magnetic quantum number $M$
belongs to ${\mathcal H}(S)$ with $S$ attaining its smallest
value $S=|M|$.
\item If $N\!\cdot\!s$ is integer, then the ground state is non-degenerate.
\item If $N\!\cdot\!s$ is half integer, then the ground state is fourfold degenerate.
\item If $s$ is integer or $N\!\cdot\!s$ even, then the shift
quantum number is $k=0$.
\item If $s$ is half integer and $N\!\cdot\!s$ odd, then the shift
quantum number turns out to be $k=N/2$.
\item If $N\!\cdot\!s$ is half integer, then $k=\lfloor(N+1)/4\rfloor$
and $k=N-\lfloor(N+1)/4\rfloor$ is found.
$\lfloor(N+1)/4\rfloor$ symbolizes the
greatest integer less or equal to $(N+1)/4$.
\item Non-degenerate ground states are also eigenstates of the
spin flip operator $\op{C}$. For half integer $s$ and
necessarily even $N$ we find that $\pi=-1$ if $N/2$ is odd and
$\pi=+1$ if $N/2$ is even. The situation is more complicated for
integer $s$. Here rows of alternating $\pi$ for odd $s$ change
with rows of $\pi=+1$ for even $s$. This behaviour was also
checked for $s=3$ and $s=4$, but not displayed in the table.
\item Non-degenerate ground states with $k\ne 0$ or $\pi\ne -1$
overlap with all product states of the Hilbert subspace with
$M=0$. For ground states with $k=0$ and $\pi=-1$ those product
states, which remain in the same cycle after application of the
spin flip, have coefficients zero.
\end{enumerate}

\section{Proofs and suggestions}

\begin{figure}[t]
\begin{center}
\epsfig{file=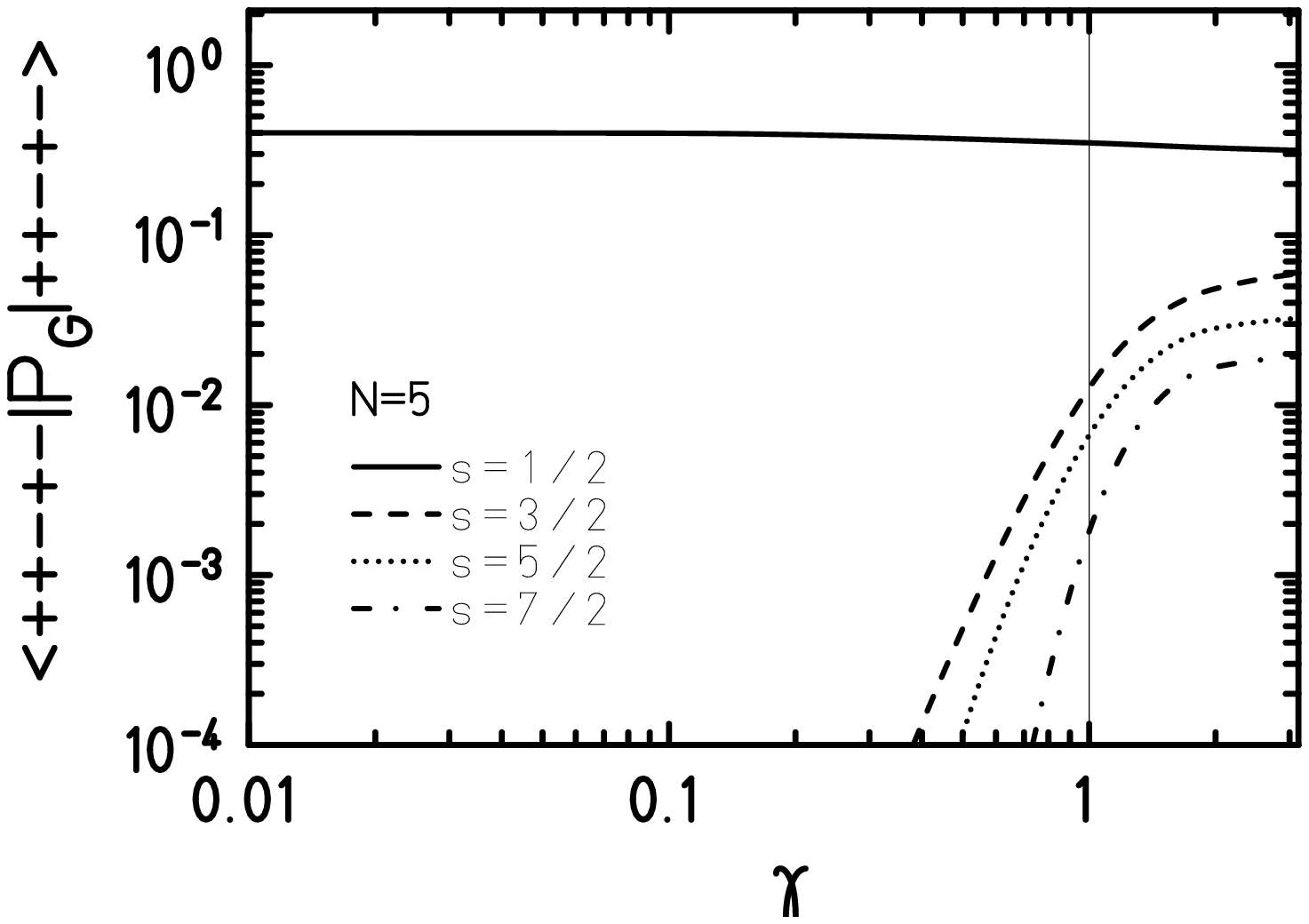,width=90mm}
\vspace*{1mm}
\caption[]{Contribution of the state
$\ket{++-+-}$ to the ground states of the hamiltonian
$\op{H}(\gamma)$, Eq. \fmref{E-2-12}.}
\label{F-3-0}
\end{center} 
\end{figure}

\begin{figure}[t]
\begin{center}
\epsfig{file=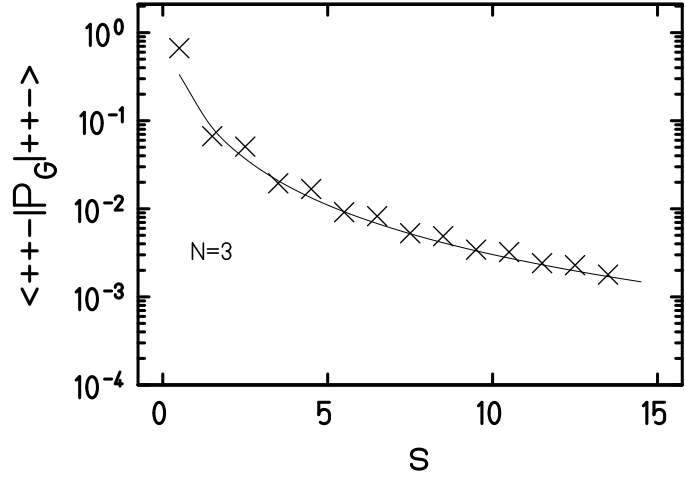,width=90mm}
\vspace*{1mm}
\caption[]{Contribution of the state $\ket{++-}$ to the ground
states of the hamiltonian $\op{H}(\gamma=1)$ (symbols). The
solid line represents the inverse of the dimension of the Hilbert
subspace with $M=1/2$.}
\label{F-3-4}
\end{center} 
\end{figure}

Although some properties, especially the first one, appear
natural, we cannot prove all of them rigorously and generally,
but we can prove special cases and provide some general
arguments:
\begin{itemize}
\item For $N\le 4$ one can evaluate the eigenvalues and
eigenstates of \fmref{E-2-1} analytically, because the Hamilton
operator can be drastically simplified
%--------------------------------------------------------
\begin{eqnarray}
\label{E-2-5}
N=2&:&
\op{H}
=
-
2\,J\,
\left(
\op{\vek{S}}^2 - \op{\vek{s}}_{1}^2 -\op{\vek{s}}_{2}^2
\right)
\ , \
\\
\label{E-2-6}
N=3&:&
\op{H}
=
-
J\,
\left(
\op{\vek{S}}^2 - \op{\vek{s}}_{1}^2 -\op{\vek{s}}_{2}^2 -\op{\vek{s}}_{3}^2
\right)
\ , \
\\
\label{E-2-7}
N=4&:&
\op{H}
=
-
J\,
\left(
\op{\vek{S}}^2 - \op{\vek{S}}_{13}^2 -\op{\vek{S}}_{24}^2
\right)
\ , \
\op{\vek{S}}_{13}=\op{\vek{s}}(1)+\op{\vek{s}}(3)
\ , \
\op{\vek{S}}_{24}=\op{\vek{s}}(2)+\op{\vek{s}}(4)
\ .
\end{eqnarray}
%--------------------------------------------------------
One realizes that the energy eigenvalues depend monotonically on
the total spin quantum number $S$ \cite{BeG90,MSL99}, so that
the AF ground state must belong to the minimal total spin
$S=\SMin$ which is either $0$ or $1/2$ with the respective
magnetic quantum numbers $M$.
\item For $N\le 4$ one can in addition explain the degeneracy of the
ground state exactly using angular momentum coupling.  It is
obvious for $N=2$ that one can couple the two spins to a total
spin running from $S=0$ to $S=2s$ with the Hilbert space of $S=0$
having dimension one. For $N=3$ one can couple the first two
spins to integers $S=0,\dots 2s$. Then for $s$ being half
integer there are two possibilities to couple to $S=1/2$ which
results in a fourfold degenerate ground state. If $s$ is
integer, one is left with only one possibility to couple to zero
giving a non-degenerate ground state.  The situation is more
involved for $N=4$. Here one needs to consider that all spin
operators $\op{\vek{S}}^2$, $\op{\vek{S}}_{13}^2$ and
$\op{\vek{S}}_{24}^2$ commute, therefore the lowest energy is
only given by the coupling where $S=0$ and $S_{13}=S_{24}=2s$,
which results in a non-degenerate ground state.
\item Non-degenerate eigenstates of $\op{H}$ remain invariant
under all unitary transformations $\op{U}$ that commute
with $\op{H}$. In the case of rotational symmetry 
this implies that the eigenstate spans a one-dimensional
irreducible representation of the rotation group, hence belongs
to $S=0$. In the case of the $k\leftrightarrow N-k$ symmetry one
concludes by an analogous argument that $k=0$ or $k=N/2$ (only
if $N$ even). Because of the spin flip symmetry non-degenerate
ground states are also eigenstates of $\op{C}$.
\item Using results of Peierls and Marshall \cite{Mar55} 
and Lieb, Schultz and Mattis \cite{LSM61,LiM62} one can
moreover prove that for even numbers $N$ of spin sites the
ground state is non-degenerate and therefore must have $S=0$. 
The proof rests on the fact that the coefficients
$c({\bold m})$ of the ground state $\ket{\Psi_0}$
with respect to the product basis
%--------------------------------------------------------
\begin{eqnarray}
\label{E-2-8}
\ket{\Psi_0}
=
\sum_{{\bold m}}
c({\bold m})\,
\ket{{\bold m}}
%\ \mbox{with}\ 
%\sum_{i=1}^{N}m_i=0
\ ,
\end{eqnarray}
%--------------------------------------------------------
with $\sum_{i=1}^{N}m_i=0$ for the ground state, 
possess a specific sign property, namely
%--------------------------------------------------------
\begin{eqnarray}
\label{E-2-9}
c({\bold m})
=
(-1)^{\left(\frac{Ns}{2}-\sum_{i=1}^{N/2} m_{2i}\right)}
a({\bold m})
\end{eqnarray}
%--------------------------------------------------------
with all $a({\bold m})$ proven to be non-zero, real
and of equal sign.  Expressing the ground state energy
$\bra{\Psi_0}\op{H}\ket{\Psi_0}$ in terms of the coefficients
$a(m_1, m_2, \dots, m_N)$ one also realizes that only one set of
coefficients can minimize the energy because having two sets one
could create a new one as the set of differences which should
belong to the same eigenspace but violates \fmref{E-2-9}.

The sketched property and its proof hold more generally in
any Hilbert subspace with constant $M$.
For rings of an odd number of spins a sign rule cannot be
established because the Hamilton operator connects basis
states $\ket{{\bold m}}$ such, that one returns to the starting
state after an odd number of steps. The simplest example for
$N=3$, $s=1/2$ reads: $\ket{++-}\rightarrow\ket{+-+}\rightarrow\ket{-++}\rightarrow\ket{++-}$.
\item A useful byproduct of the sign rule \fmref{E-2-9} is that
it also explains the sequence of $k$-values for even
$N$. Consider the action of the cyclic shift operator on the
basis states \fmref{E-2-3}. The change in sign of the
coefficient, whose absolute value is not altered, then is
%--------------------------------------------------------
\begin{eqnarray}
\label{E-2-10}
\frac{c(m_1, \dots, m_{N-1}, m_N)}
     {c(m_N, m_1, \dots, m_{N-1})}
=
(-1)^{\left(\sum_{i=1}^{N/2} m_{2i}-\sum_{i=1}^{N/2} m_{2i-1}\right)}
=
(-1)^{\left( N s \right)}
\ .
\end{eqnarray}
%--------------------------------------------------------
Thus for odd $N\!\cdot\!s$ we find $k=N/2$, whereas even
$N\!\cdot\!s$ implies $k=0$.
\item For even $N$ the ground state has non-zero
components with respect to product states of the form
$\ket{m,-m,m,-m,\dots}, m\ne 0$. This explains the correlation
between $k$ and $\pi$, namely $k=0 \leftrightarrow \pi=+$ and
$k=N/2 \leftrightarrow \pi=-$, because cyclic shift and spin
flip operator have the same effect on these product states.  One
may also observe that an alternating product state
$\ket{m,-m,m,-m,\dots}$ generates a cycle of dimension $2$
(cf. \cite{BSS99}) which in turn implies the selection rule
$k=0$ or $k=N/2$.
\item If $N\!\cdot\!s$ is half integer, then the ground state, which has
a finite total magnetic quantum number $M$, must be at least
twofold degenerate because of the spin flip symmetry $\forall x:
m_x \rightarrow -m_x$, i.e. $M \rightarrow -M$.  
If one could show, that $k$ could not be zero, one would explain
the fourfold degeneracy. It is, however,
not possible to derive the fourfold degeneracy for such cases
using only symmetry arguments since the spectrum contains also
twofold degenerate energy eigenvalues.
\item \changed{For $s=1/2$ the fourfold degeneracy and the special
$k$-values can be explained using the Bethe ansatz for odd $N$
\cite{Kar94}.} 
\end{itemize}
At the end of this section we would like to suggest an
explanation for the special $k$-values found for half integer
$N\!\cdot\!s$. 
It is sufficient to restrict the investigations
to the Hilbert subspace ${\mathcal H}(M=1/2)$. 
Our arguments rest on the observation that the cycle generated
by the special basis state
%--------------------------------------------------------
\begin{eqnarray}
\label{E-2-11}
\ket{{\bold m}_0}
:=
\ket{\half\ , \half\ , -\half\ , \half\ , -\half\ , \half\ , -\half\cdots}
=
\ket{+ + - + - + - \cdots}
\ ,
\end{eqnarray}
%--------------------------------------------------------
which is maximally alternating but minimally undulating,
contributes to the ground state. For $s=1/2$ this seems to be
rather natural, because this special cycle constitutes the
ground state of the Ising model (\eqref{E-2-12}, $\gamma=0$) and
it is very likely that it also contributes to the ground state
of the respective Heisenberg hamiltonian (\eqref{E-2-12},
$\gamma=1$)
%--------------------------------------------------------
\begin{eqnarray}
\label{E-2-12}
\op{H}(\gamma)
&=&
-
2 J
\sum_{x}\;
\left\{
\op{s}^3(x) \op{s}^3(x+1)
+
\frac{\gamma}{2}
\left[
\op{s}^+(x) \op{s}^-(x+1)
+
\op{s}^-(x) \op{s}^+(x+1)
\right]
\right\}
\ .
\end{eqnarray}
%--------------------------------------------------------
This behaviour is demonstrated by \figref{F-3-0} for a ring of
five spins. The figure also shows that this cycle is a ground
state component for higher spin quantum numbers, too; which is
found consistently for all investigated systems. Figure
\figref{F-3-4} suggests for the example of three spins that the
weight of the special cycle is about the inverse of the
dimension of the Hilbert subspace ${\mathcal H}(M=1/2)$.
Considering the expectation value of the Hamilton
operator in a ground state $\ket{\Psi_0; k}$ with translational
quantum number $k$ 
and especially the contribution of the cycle generated
by $\ket{{\bold m}_0}$ one finds
%--------------------------------------------------------
\begin{eqnarray}
\label{E-2-13}
\bra{\Psi_0; k}\op{H}\ket{\Psi_0; k}
&=&
-4 J\,
\left(c({\bold m}_0)\right)^2\, N\,
\cos\left(
2\,\frac{2 \pi k}{N}
\right)
\\
&& + \mbox{remaining terms}
\nonumber
\ .
\end{eqnarray}
%--------------------------------------------------------
The argument $2\,\frac{2 \pi k}{N}$ of the cosine function stems
from the fact that
%--------------------------------------------------------
\begin{eqnarray}
\label{E-2-14}
\bra{{\bold m}_0}\op{H}\,\op{T}^{\nu}\ket{{\bold m}_0}
\propto
\left\{
\delta_{\nu,2}
+
\delta_{\nu,N-2}
\right\}
\ .
\end{eqnarray}
%--------------------------------------------------------
Assuming that minimization of the first term of the r.h.s. of
\fmref{E-2-13} cannot be counteracted enough by the
$k$-dependence of the remaining terms, one understands that in
the ground state $k$ must be as close as possible to
$N/4$ or $3N/4$, hence $k=\lfloor(N+1)/4\rfloor$ or
$N-\lfloor(N+1)/4\rfloor$.  

It will be a subject of further research whether this $k$
selection rule holds outside the range of $N$ and $s$ studied
so far.

\section{Degeneracy and frustration effects}

Our results suggest a rather different behaviour for spin rings
with non-degenerate and degenerate ground states. 

\begin{figure}[t]
\begin{center}
\epsfig{file=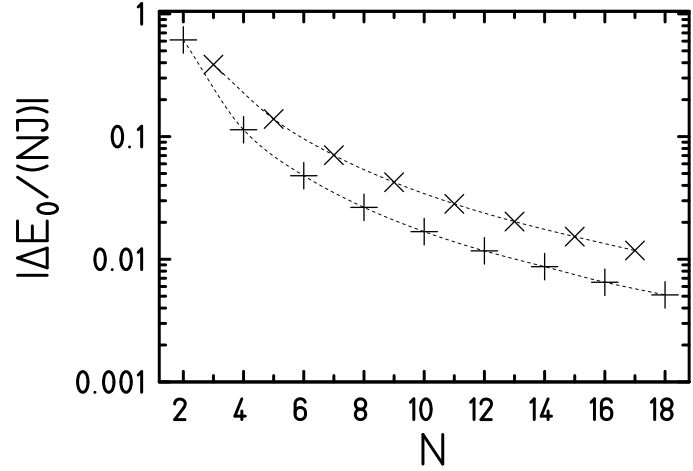,width=90mm}
\vspace*{1mm}
\caption[]{Deviation of ground state energies (symbols) for
antiferromagnetic coupled Heisenberg rings ($s=1/2$) from the
large $N$ limit of Bethe and Hulth\'{e}n
\cite{Bet31,Hul38}. Plus symbols are used for even $N$, crosses
for odd $N$.}
\label{F-3-1}
\end{center} 
\end{figure}

The first example is given by the ground state energies of
spin-$\half$-rings which converge to the Bethe-Hulth\'{e}n
limit of $E/(NJ)=2\ln(2)-1/2$ \cite{Bet31,Hul38}. As
\figref{F-3-1} shows this convergence is faster for even numbers
of spins than for odd numbers. The latter systems
are also frustrated.

\begin{figure}[t]
\begin{center}
\epsfig{file=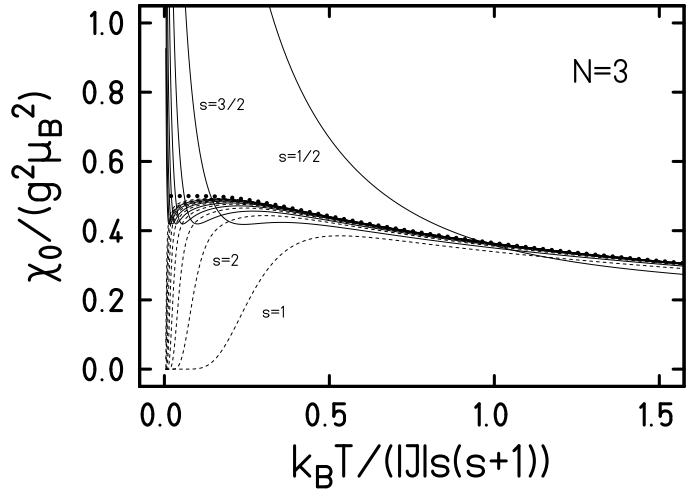,width=90mm}
\vspace*{1mm}
\caption[]{Zero-field susceptibility for $N=3$ and various
$s=1/2,1,\dots,9$. The solid lines show the result for half
integer spins, the dashed lines for integer spin quantum
numbers. The classical result is given by the thick dotted line.}
\label{F-3-2}
\end{center} 
\end{figure}

In the second example effects of the ground state degeneracy on
the zero-field magnetic susceptibility are discussed. To this
end we introduce the interaction of all magnetic moments with
the homogeneous magnetic field $B$ (Zeeman term) into the
Hamilton operator
%--------------------------------------------------------
\begin{eqnarray}
\label{E-3-1}
\op{H}_{\!M}
&=&
-
2\,J\,
\sum_{x=1}^N\;
\op{\vek{s}}(x) \cdot \op{\vek{s}}(x+1)
+
g \mu_B B \op{S}^3
=
\op{H}
+
g \mu_B B \op{S}^3
\ .
\end{eqnarray}
%--------------------------------------------------------
The magnetisation ${\mathcal M}$ is defined as
%--------------------------------------------------------
\begin{eqnarray}
\label{E-3-2}
{\mathcal M}
&=&
\frac{1}{Z}\,
\tr\left\{-g \mu_B \op{S}^3 \mbox{e}^{-\beta\op{H}_{\!M}}\right\}
\ ,\quad
Z
=
\tr\left\{\mbox{e}^{-\beta\op{H}_{\!M}}\right\}
\end{eqnarray}
%--------------------------------------------------------
and its derivative with respect to the magnetic field $B$
results in the zero-field susceptibility
%--------------------------------------------------------
\begin{eqnarray}
\label{E-3-3}
\chi_0
&=&
\left(\pp{{\mathcal M}}{B}\right)_{B=0}
=
g^2 \mu_B^2 \beta
\left(
\frac{1}{Z}\,
\tr\left\{\left(\op{S}^3\right)^2
\mbox{e}^{-\beta\op{H}_{\!M}}\right\}
\right)_{B=0}
\ .
\end{eqnarray}
%--------------------------------------------------------
Figure \xref{F-3-2} displays the zero-field susceptibility for
$N=3$ and various $s=1/2,1,\dots,9$. For small temperatures the
susceptibility diverges for half integer spins (solid lines),
whereas it drops to zero for integer spin quantum numbers
(dashed lines), see also \cite{BoF64}. The different behaviour
is easy to understand. In the degenerate case two eigenstates
have a total magnetic quantum number $M=\half$ and the other two
have $M=-\half$. The slightest magnetic field suffices to split
the degeneracy in $M$ and thus results in a finite magnetic
moment even at $T=0$, because the new twofold degenerate ground
state has $M=-\half$. Therefore in general, for half integer
spin quantum numbers and odd $N$
%--------------------------------------------------------
\begin{eqnarray}
\label{E-3-4}
\frac{\chi_0 k_B T}{g^2 \mu_B^2}
\stackrel{T\rightarrow 0}{\longrightarrow}
\frac{1}{4}
\ .
\end{eqnarray}
%--------------------------------------------------------
In the non-degenerate cases the magnetisation at small $B$ can
only grow by thermally populating excited states with
non-vanishing magnetic moment.

The classical result \cite{CLA99} is given by the thick dotted
line. It neither goes to infinity nor to zero but to $1/2$ in
the used units.

\changed{Our general results on the ground state degeneracy thus
enable us to describe the zero-field susceptibility at low
temperatures qualitatively, for example we know, consequently,
that $\chi_0$ drops to zero for a ring of $N=11$ and $s=2$ which
hardly could be calculated.}

\begin{figure}[t]
\begin{center}
\epsfig{file=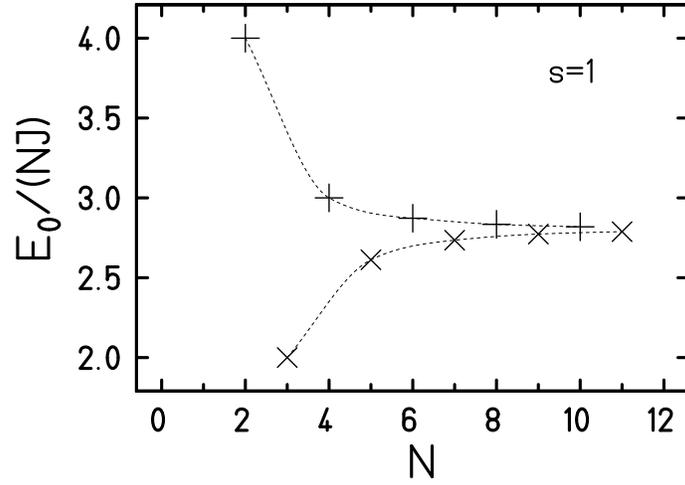,width=90mm}
\vspace*{1mm}
\caption[]{Ground state energies for antiferromagnetically coupled
Heisenberg rings ($s=1$). Plus symbols are used for even
$N$, crosses for odd $N$. Frustration expresses itself through
a weaker binding for odd $N$.}
\label{F-3-3}
\end{center} 
\end{figure}

At this point some words regarding frustration might be
\changed{also} in order. In classical physics an
antiferromagnetic spin system is called frustrated if not all
spins can be paired. Then the ground state possesses a
non-trivial degeneracy. An example is the spin triangle in the
Heisenberg model which shows a non-trivial twofold degeneracy,
see for instance \cite{Cho86,Kaw98}.
These considerations regarding the interplay of ground state
degeneracy and frustration are not naively applicable to
Heisenberg rings with isotropic next neighbor interaction,
although it is appealing and one could conjecture it from
the first example. For quantum spin rings we find that rings with
an odd number of integer spins have a non-degenerate ground
state. That they are nevertheless frustrated can be seen
comparing their ground state energy per spin with their even
neighbours, see \figref{F-3-3}.  The frustration expresses itself
through a weaker binding for odd $N$.

%%%%%%%%%%%%%%%%%%%%%%%%%%%%%%%%%%%%%%%%%%%%%%%%%%%%%%%%%%%%%%%%%%%%%%%%
\section*{Acknowledgments}
The authors would like to thank M.~Luban (Ames), D.~Mentrup and
J.~Richter and his group (Magdeburg) for stimulating discussions.
%%%%%%%%%%%%%%%%%%%%%%%%%%%%%%%%%%%%%%%%%%%%%%%%%%%%%%%%%%%%%%%%%%%%%%%%

%-----------------------------------------------------------------------
\begin{table}[t]
\begin{center}
\begin{tabular}{|cc||r|r|r|r|r|r|r|r|r|l|}
\hline
&&\multicolumn{9}{c|}{$N$}&\\
&& 2&3&4&5&6&7&8&9&10&\\
\hline\hline
\multirow{18}{0mm}{$s$}
&             & 1.5 & 0.5  & 1 & 0.747 & 0.934 & 0.816 
& 0.913 & 0.844 & 0.903 & $E/(NJ)$\\
&$\frac{1}{2}$& 1   & 4    & 1 & 4        & 1        & 4        
& 1        & 4        & 1 & $deg$\\
&             & 1   & 1, 2 & 0 & 1, 4     & 3        & 2, 5
& 0        & 2, 7     & 5 & $k$\\
&             & $-$ &      &$+$&          & $-$      &     
& $+$      &          &$-$& $\pi$\\
\cline{2-12}
&   & 4 & 2 & 3 & 2.612 &2.872&2.735&2.834&2.773&2.819& $E/(NJ)$\\
&$1$& 1 & 1 & 1 & 1     & 1   & 1   & 1   & 1   & 1   &  $deg$\\
&   & 0 & 0 & 0 & 0     & 0   & 0   & 0   & 0   & 0   & $k$\\
&   &$+$&$-$&$+$& $-$   & $+$ & $-$ & $+$ & $-$ & $+$ & $\pi$\\
\cline{2-12}
&             & 7.5 & 3.5 & 6 &4.973&5.798&5.338&5.732&5.477&$5.704^{\dagger\dagger}$& $E/(NJ)$\\
&$\frac{3}{2}$& 1   & 4   & 1 & 4   & 1   & 4   & 1   & 4   & 1 &  $deg$\\
&             & 1   & 1, 2& 0 & 1, 4& 3   & 2, 5& 0   & 2, 7& 5 & $k$\\
&             &$-$  &     &$+$&     & $-$ &     & $+$ &     &     & $\pi$\\
\cline{2-12}
&   & 12 & 6 & 10 &8.456&9.722&9.045&9.630&$9.263^{\dagger\dagger}$& & $E/(NJ)$\\
&$2$& 1  & 1 & 1  & 1   & 1   & 1   & 1   & 1 & &  $deg$\\
&   & 0  & 0 & 0  & 0   & 0   & 0   & 0   & 0 & & $k$\\
&   &$+$ &$+$&$+$ & $+$ & $+$ & $+$ &     &     &     & $\pi$\\
\cline{2-12}
&             & 17.5 & 8.5 & 15 &12.434&14.645&13.451&$14.528^{\dagger}$&& & $E/(NJ)$\\
&$\frac{5}{2}$& 1    & 4   & 1  & 4    & 1    & 4    & 1    && &  $deg$\\
&             & 1    & 1, 2& 0  & 1,4  & 3    & 2, 5 & 0    && & $k$\\
&             & $-$  &     & $+$&      & $-$  &      &      && & $\pi$\\
\hline
\end{tabular}
\vspace*{5mm}
\begin{tabular}{|cc||r|r|r|r|r|r|r|r|l|}
\hline
&&\multicolumn{8}{c|}{$N$}&\\
&& 11&12&13&14&15&16&17&18&\\
\hline\hline
\multirow{3}{0mm}{$s$}
&             &0.858&0.898&0.866&0.895&0.871&0.893&0.874&0.891& $E/(NJ)$\\
&$\frac{1}{2}$& 4   & 1   & 4   & 1   & 4   & 1   & 4   & 1   & $deg$\\
&             & 3, 8& 0   &3, 10& 7   &4, 11& 0   &4, 13& 9   & $k$\\
\hline
\end{tabular}
\vspace*{5mm}
\end{center}
\caption{Properties of the AF Heisenberg ground state, energy
$E$, degeneracy $deg$, shift quantum number $k$ and spin
flip parity $\pi$. Values for the empty fields could not be
computed in reasonable times.
$\dagger$ -- O.~Waldmann, private communication.
$\dagger\dagger$ -- projection method \cite{Man91}.
}\label{T-2-1}
\end{table}
%----------------------------------------------------------------------- 

%===================    figure   =================================
%===================    figure   =================================

%===================    figure   =================================
%===================    figure   =================================

%===================    figure   =================================
%===================    figure   =================================

%===================    figure   =================================
%===================    figure   =================================

%===================    figure   =================================
%===================    figure   =================================

\end{document}